\documentclass[a4paper,12pt]{amsart}

\usepackage{amssymb,amsmath,dsfont,mathtools,framed,bm,esint}
\numberwithin{equation}{section}

\newcommand{\CC}{\mathbb{C}}
\newcommand{\RR}{\mathbb{R}}
\newcommand{\abs}[1]{\left\vert #1\right\vert}

\theoremstyle{remark}
\newtheorem{remark}{Remark}

\usepackage[pdfdisplaydoctitle=true,
            colorlinks=true,
            urlcolor=blue,
            citecolor=blue,
            linkcolor=blue,
            bookmarksnumbered=true]{hyperref}
            
\title[Solitons and energy dependent spectral problems]{Solitons of shallow-water models from energy-dependent spectral problems}
\author{Jack Haberlin}
\address[Jack Haberlin]{Department of Science, Waterford Institute of Technology, Cork Road, Waterford, Ireland}
\email{20063228@mail.wit.ie}

\author{Tony Lyons}
\address[Tony Lyons]{Department of Computing and Mathematics, Waterford Institute of Technology, Cork Road, Waterford, Ireland}
\email[Corresponding author]{tlyons@wit.ie}

\begin{document}
\maketitle
\begin{abstract}
The current work  investigates the soliton solutions of the Kaup-Boussinesq equation using the Inverse Scattering Transform method. We outline the construction of the Riemann-Hilbert problem for a  pair energy-dependent spectral problems for the system, which we then use to  construct the solution of this hydrodynamic system.
\end{abstract}

\section{Introduction}
Many nonlinear systems which we use to model natural phenomena allow for soliton solutions. In particular, the mathematical theory of hydrodyanmics provides a rich environment for the study of soliton theory. Perhaps the simplest nonlinear hydrodyanmic model is the KdV equation \cite{KdV1895}, derived as a model for inviscid shallow-water waves. While KdV was initially known to possess localised wave solutions we come to expect of solitons, the formal construction of these solutions by means of the Inverse Scattering Transform (IST), was achieved relatively recently in the seminal work \cite{GGKM1967}. Following on this, the Lax formalism \cite{Lax1968} ensured the IST could be used to construct the soliton solutions of many other nonlinear systems. For an  overview of many  modern and classical results of soliton theory, which we use throughout, the reader is referred to the following monographs \cite{AS1981,Con2011,DJ1989, FT1987, GVY2008, Joh1997, New1985,NMPZ1984,PS2010}

In the work \cite{IL2012} the  Lax pair given by
\begin{equation}\label{eq1.1}
\begin{cases}
\Psi_{xx} = \left(-\lambda^{2} + \lambda u + \frac{\kappa}{2}u^{2} + \eta\right)\Psi,\\
\Psi_{t}= -\left(\lambda+\frac{1}{2}u\right)\Psi_{x}+\frac{1}{4}u_{x}\Psi,
\end{cases}
\end{equation}
is introduced as an \emph{energy dependent} spectral problem, where $\lambda$ is the spectral parameter, and $\kappa$ is a physical constant. In analogy with the Schr\"{o}dinger equation, where the spectral parameter labels the energy levels of the system, the potetial in equation \eqref{eq1.1} clearly depends on this spectral parameter, and so the moniker ``energy depdendent'' spectral problem is used to describe this Lax pair.  The potentials $u$ and $\eta$ are understood to depend on $x$ and $t$, while the spectral functions $\Psi$ are understood to be dependent on the variables $x$, $t$  and the spectral parameter $\lambda$. Other systems obtained from energy dependent spectral problems, among them the Kaup-Boussinesq and two-component Camassa-Holm equations, are studied in the following works \cite{AF1989,AFL1991,BPZ2001,CI2008,HI2011, Iva2009}.

From the compatibility condition  $\Psi_{xxt}=\Psi_{txx}$, we obtain a family of two-component integrable systems parameterised by $\kappa$, and  possessing terms with cubic nonlinearities, see \cite{IL2012}. Nevertheless, the choice $\kappa=-\frac{1}{2}$ yileds a particulary interesting system possessing only quadratic nonlinearities:
\begin{equation}{\label{KB}}\tag{KB}
\begin{cases}
u_{t}+ \eta_{x} + uu_{x}=0\\
\eta_{t}- \frac{1}{4}u_{xxx}+(u\eta)_x=0,
\end{cases}
\end{equation}
which was introduced as an approximate model of shallow-water waves in \cite{Kau1975}. In this case the potential $u$ is interpreted as the fluid velocity while the potential $\eta$ corresponds to the surface elevation of the fluid. The reader is referred to \cite{Iva2009} where a derivation of the system \eqref{KB} is presented as a model of shallow water waves in the presence of a linear shear current, while the system also appears as a hydrodynamic model in the works \cite{EGP2001,Whi2011,DJ1989}.

The work \cite{Kau1975} also outlined a method of solving the system \eqref{KB} by means of the Inverse Scattering Transform via an appropriate Lax pair, while the Hamiltonian hierarchy is investigated in the work \cite{Pav2002}. The solutions of the system \eqref{KB} are also studied in the works \cite{EGK2005,KKU2003,MY1979}, while spectral problems with energy-dependent potentials were also studied in the works \cite{Jau1972, JJ1972, JJ1976, JJ1981, SS1996, LSS2007}.

In the current work we will utilise the IST for the system \eqref{KB}, developed in the work \cite{IL2012},  where an explcit topological solution of the system was presented. In this paper we  will construct an explicit soliton solution for the system \eqref{KB}. While initially it appears two discrete speactral data points $\{\lambda_1,\lambda_2\}$ are required to construct this soliton solution, the reduction symmetries of the spectral problem to be used will ensure these data are related by $\lambda_1=-\bar{\lambda}_2$. It will be found that the solution $\eta$ is symmetric about $x=0$, while $u$ is anti-symmetric about the same axis, in line with the solution obtained in \cite{IL2012}.

\section{Preliminaries}
\subsection{The conjugate spectral problems}
To implement the inverse scattering transform for the \eqref{KB}, the spectral problem in \eqref{eq1.1} is extended to a pair of conjugate spectral problems given by
\begin{equation}\label{eq2.1}
\begin{cases}
\Psi_{xx}(x,\lambda,\sigma) = \left(-\lambda^{2} + \sigma\lambda u + w\right)\Psi(x,\lambda\sigma),\\
 w = \frac{\kappa^2}{2}u + \eta,\\
 \sigma =\pm 1.
\end{cases}
\end{equation}
We refer the reader to \cite{IL2012} for a full overwiew of the IST of this family of spectral problems. It is assumed in this work that the potentials satisfy  $(u,\eta)\in\mathcal{S}(\mathbb{R})^2$, where $\mathcal{S}(\mathbb{R})$ denotes that Schwartz space of rapidly decreasing funcitons on $\mathbb{R}$. Owing to this rapid decay of the potentials, we find that the solutions of the conjugate spectral porblems \eqref{eq2.1} exhibit the following asymptotic behaviour:
\begin{equation}\label{eq2.2}
\begin{split}
\psi_{1}(x,\lambda,\sigma)\to e^{-i\lambda x}\qquad \psi_{2}(x,\lambda,\sigma)\to e^{i\lambda x}\quad\text{ as }x\to+\infty\\
\phi_{1}(x,\lambda,\sigma)\to e^{-i\lambda x}\qquad \phi_{2}(x,\lambda,\sigma)\to e^{i\lambda x}\quad\text{ as }x\to-\infty,
\end{split}
\end{equation}
which we use to define the Jost solutions of the spectral problem. Here and throuhout the time-dependence of the Jost solutions and various other functions shall be left implicit for convenience.

The invariance of equation \eqref{eq2.1} under the involution $(\lambda,\sigma)\to(-\lambda,-\sigma)$ in conjunction with the asymptotic behavior of the $u$ and $w$, leads us to deduce
\begin{equation}\label{eq2.3}
    \psi_1(x,\lambda,\sigma)=\psi_2(x,-\lambda,-\sigma)\quad\phi_1(x,\lambda,\sigma)=\phi_2(x,-\lambda,-\sigma).
\end{equation}
Moreover, it is immediately clear from \eqref{eq2.1} that if $\psi(x,\lambda,\sigma)$ is a solution of the spectral problem, then so too is $\bar{\psi}(x,-\bar{\lambda},-\sigma)$. Thus upon comparing the asymptotic behaviour of the Jost solutions,we find
\begin{equation}\label{eq2.4}
    \psi_{1}(x,\lambda,\sigma)=\bar{\psi}_{2}(x,-\bar{\lambda},-\sigma),\quad\phi_{1}(x,\lambda,\sigma)=\bar{\phi}_{2}(x,-\bar{\lambda},-\sigma),
\end{equation}
thereby allowing us to construct two bases of solutions using the Jost functions $\psi(x,\lambda,\sigma) = \psi_1(x,\lambda,\sigma)$ and $\phi(x,\lambda,\sigma) = \phi_1(x,\lambda,\sigma)$
along with their conjugate partners.

Introducing the charge
\begin{equation}\label{eq2.5}
\begin{split}
&\omega_+ :=\frac{1}{2}\int_{x}^{\infty}u(s,t)\,ds,\\
\end{split}
\end{equation}
it is easily deduced from the system \eqref{KB} that $\alpha_1=\omega_{+}(-\infty,t)$ is conserved, owing to the asymptotic decay of the potential.
It proves helpful to define the following augmented Jost functions
\begin{equation}\label{eq2.6}
\underline{\psi}(x,\lambda,\sigma) = \psi(x,\lambda,\sigma)e^{i(\lambda x+\sigma\omega_+)},\quad
\underline{\phi}(x,\lambda,\sigma) = \phi(x,\lambda,\sigma)e^{i(\lambda x-\sigma\omega_+)},
\end{equation}
and it may be shown that $\underline{\psi}(x,\lambda,\sigma)$ is analytic for $\lambda \in \mathbb{C_-}$ while
$\underline{\phi}(x,\lambda,\sigma)$ is analytic for $\lambda \in \mathbb{C_+}$ (see \cite{IL2012} \S4), where $\mathbb{C}_{\pm}$ denote the upper and lower half of the complex plane.

\subsection{The Riemann-Hilbert problem}
Since $\psi$ and its conjugate partner constitute a basis of solutions for the spectral problem in \eqref{eq2.1}, we may write
\begin{equation}\label{eq2.7}
\phi(x,\lambda,\sigma) = a(\lambda,\sigma)\psi(x,\lambda,\sigma)+ b(\lambda,\sigma)\bar{\psi}(x,\lambda,\sigma),
\end{equation}
where $a$ and $b$ denote the scattering coefficients associated with the spectral problem. Written in terms of the augmented Jost functions, this scattering relation becomes
\begin{equation}\label{eq2.8}
\frac{\underline{\phi}(x,\lambda,\sigma)e^{i\sigma\alpha_{1}}}{a(\lambda,\sigma)}=\underline{\psi}(x,\lambda,\sigma)
+r(\lambda,\sigma)\underline{\bar{\psi}}(x,\lambda,\sigma)e^{2i(\lambda x+\sigma\omega_{+})}\quad \lambda\in\mathbb{R},
\end{equation}
where we define the reflection coefficient $r(\lambda,\sigma)=\frac{b(\lambda,\sigma)}{a(\lambda,\sigma)}$.

It was shown in  \cite{IL2012} that the scattering coefficient $a(\lambda,\sigma)$ may be analytically continued throughout $\lambda\in\CC_+,$
while it was also assumed  that this scattering coefficient has a finite number of simple zeros at $\lambda_n\in \CC_+$, where $1\leq n\leq N$.
These simple zeros constitute the the discrete spectrum of the spectral
problem \eqref{eq2.1}.  We observe that the right-hand side of equation \eqref{eq2.8} is a functions analytic throughout $\lambda\in\CC_-$ while the left-hand side is analytic throughout $\CC_+$ except for a finite number of simple poles, hence the equation represents a Riemann-Hilbert problem with a jump along the real line.

\begin{figure}
\centering
\includegraphics[width=\textwidth]{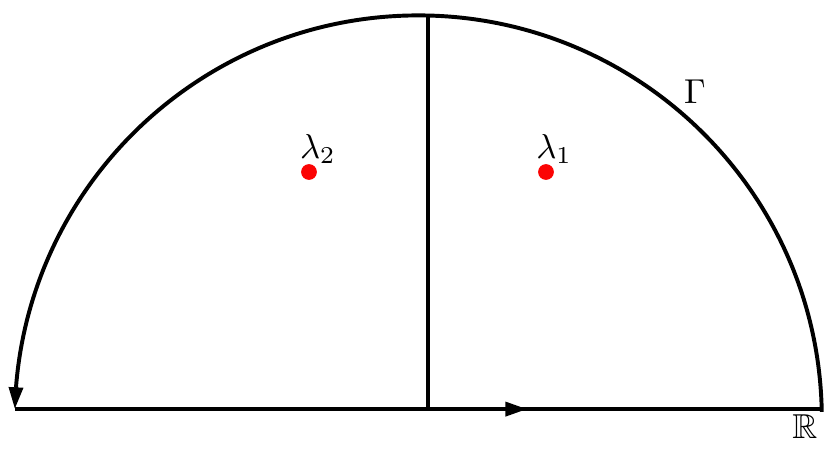}
\caption{The contour $C_+$ enclosing the discrete spectrum $\lambda_1$ and $\lambda_2$.}
\label{RH}
\end{figure}

Choosing $\lambda\in\CC_-$ arbitrarily, we integrating \eqref{eq2.8} around the contour $C_+$ displayed in Figure \ref{RH}, and upon
applying the residue theorem we find that
\begin{equation}\label{eq2.9}
\frac{1}{2\pi i}\ointctrclockwise_{C_{+}}\frac{\underline{\phi}(\lambda',\sigma)e^{i\sigma\alpha_{1}}}{a(\lambda',\sigma)
\cdot(\lambda'-\lambda)}d\lambda'
=\sum_{n=1}^{N}\frac{\underline{\phi}^{(n)}(\sigma)e^{i\sigma\alpha_{1}}}{\dot{a}_{n}(\sigma)\cdot(\lambda_{n}-\lambda)}, \end{equation}
where we define
\[\dot{a}_{n}(\sigma)=\left(\frac{da(\lambda,\sigma)}{d\lambda}\right)_{\lambda=\lambda_n}\quad
\underline{\phi}^{(n)}(\sigma)=\underline{\phi}(\lambda_n,\sigma).\]
Additionally,  we may split the left-hand integral in \eqref{eq2.9} into a contribution along $\RR$ and a contribution from $\Gamma,$
and upon using the scattering relation \eqref{eq2.7} we find
\begin{multline}\label{eq2.10}
\sum_{n=1}^{N}\frac{\underline{\phi}^{(n)}(x,\sigma)e^{i\alpha_{1}}}{\dot{a}(\lambda_{n})\cdot(\lambda_{n}
- \lambda)}=\frac{1}{2\pi i}\int_{\Gamma_{+}}\frac{1}{\lambda^{\prime}-\lambda}d\lambda^{\prime} +
\frac{1}{2\pi i}\int_{-\infty}^{\infty}\frac{\underline{\psi}(x,\lambda',\sigma)}{\lambda^{\prime}-\lambda}d\lambda^{\prime}\\ +\frac{1}{2\pi i}\int_{-\infty}^{\infty}\frac{r(\lambda^{\prime},\sigma)e^{2i(\lambda^{\prime}x+\sigma\omega_{+})}
\underline{\bar{\psi}}(\lambda^{\prime},\sigma)}{\lambda^{\prime}-\lambda}d\lambda^{\prime}
\end{multline}
having used $\underline{\psi}(x,\lambda,\sigma)= 1$ on $\Gamma.$ In addition we may use the analyticity of $\underline{\psi}(x,\lambda,\sigma)$ in ${\CC}_-$ and integrate over the contour $-{C}_+\subset\CC_-$ to evaluate the first two terms in \eqref{eq2.10} to yield
\begin{multline}\label{eq1.15}
\underline{\psi}(x,\lambda,\sigma) = 1 -
\sum_{n=1}^{N}\frac{\underline{\phi}^{(n)}(x,\sigma)e^{i\alpha_{1}}}{\dot{a}_{n}(\sigma)(\lambda_{n}-\lambda)}
\\+\frac{1}{2\pi i}\int_{-\infty}^{\infty}\frac{r(\lambda',\sigma)e^{2i(\lambda'x+\sigma\omega_{+})}\underline{\bar{\psi}}(x,\lambda',\sigma)}{\lambda'-\lambda}d\lambda',
\end{multline}
and we refer the reader to \cite{IL2012} for further discussion.
We also note that along the discrete spectrum we have $a(\lambda_n\sigma)=0$ and so equation \eqref{eq2.7} becomes
\[\phi(x,\lambda_n,\sigma)=b_n(\sigma)\bar{ \psi}(x,\bar{\lambda}_n, \sigma).\]
Thus we find
\begin{multline}\label{eq1.16}
\underline{\psi}(x,\lambda,\sigma)=1+
i\sum_{n=1}^{N}\frac{R_{n}(\sigma)}{(\lambda-\lambda_{n})}\underline{\bar{\psi}}(\bar{\lambda}_n,\sigma)e^{2i(\lambda_{n}x+\sigma\omega_{+})} \\+\frac{1}{2\pi i} \int_{-\infty}^{\infty}\frac{r(\lambda',\sigma)}{\lambda'-\lambda}\underline{\bar{\psi}}(\lambda',\sigma)e^{2i(\lambda'x+\sigma\omega_{+})}d\lambda',
\end{multline}
where $R_{n}(\sigma)=\frac{b_n(\sigma)}{\dot{a}_n(\sigma)}$, yielding an algebraic-integral system for the augmented Jost functions $\underline{\psi}(x,\lambda,\sigma).$

\section{Symmetries and the Discrete Jost Solutions}\label{sec3}
\subsection{The scattering relations}\label{sec3.1}
Soliton solutions arise specifically when the reflection coefficient $r(\lambda,\sigma)$ vanishes on the continuous spectrum $\lambda\in\RR$. The soliton solution we seek corresponds to the case of two points on the discrete data spectrum and consequently we will construct our solution from the algebraic system
\begin{equation}\label{eq3.1}
\begin{split}
\underline{\psi}(x,\lambda,\sigma)&=1+i\sum_{n=1}^{2}\frac{{R}_n(\sigma)}{\lambda-\lambda_n}\underline{\bar{\psi}}(x,\bar{\lambda}_n,\sigma)e^{2i({\lambda}_1x+\sigma\omega_+)}\\
\underline{\bar{\psi}}(x,-\lambda,-\sigma)&=1+i\sum_{n=1}^{2}\frac{\bar{R}_n(-\sigma)}{\lambda+\bar{\lambda}_n}\underline{{\psi}}(x,\bar{\lambda}_n,-\sigma)e^{-2i(\bar{\lambda}_1x-\sigma\omega_+)}
\end{split}
\end{equation}
Clearly the first expression in equation \eqref{eq3.1} has simple poles at $\lambda=\lambda_{1}$ and $\lambda=\lambda_2$, while the second expression has poles when $\lambda=-\bar{\lambda}_1$ and $\lambda=-\bar{\lambda}_2$.  Since these are the same solution, we require that the poles match in each expression. This occurs  when either of the following conditions are satisfied:
\begin{equation}\label{eq3.2}
\begin{split}
\lambda_1=-\bar{\lambda}_1,\ \lambda_2=-\bar{\lambda}_2\quad\text{ or }\quad\lambda_2=-\bar{\lambda}_1.
\end{split}
\end{equation}
Implementing $\lambda_2 = -\bar{\lambda}_1$ and using
$\underline{\bar{\psi}}(x,-\bar{\lambda},-\sigma) = \underline{\psi}(x,\lambda,\sigma)$, the equivalence of each expression in equation \eqref{eq3.1} then requires
\begin{equation}\label{eq3.3}
\begin{split}
\frac{i\bar{R}_{1}(-\sigma)\underline{{\psi}}(x,\bar{\lambda}_1,-\sigma)}{\bar{\lambda}_{1}+\lambda}e^{-2i(\bar{\lambda}_1x-\sigma\omega_+)}
+\frac{i\bar{R}_{2}(-\sigma)\underline{{\psi}}(x,-{\lambda}_1,-\sigma)}{-{\lambda}_{1}+\lambda}e^{2i({\lambda}_1x+\sigma\omega_+)}\\
=\frac{iR_{1}(\sigma)\underline{\bar{\psi}}(x,\bar{\lambda}_1,\sigma)}{-\lambda_{1}+\lambda}e^{2i(\lambda_1x+\sigma\omega_+)}
+\frac{iR_{2}(\sigma)\underline{\bar{\psi}}(x,-\lambda_1,\sigma)}{\bar{\lambda}_{1}+\lambda}e^{2i(-\bar{\lambda}_1x+\sigma\omega_+)},
\end{split}
\end{equation}
and equating expressions with equivalent poles, we deduce
\[R_{1}(\sigma) = \bar{R}_2(-\sigma).\]
Hence we find
\begin{equation}\label{eq3.4}
\underline{\psi}(x,\lambda,\sigma) = 1 + \frac{iR_{1}(\sigma)e^{2i(\lambda_1x+\sigma\omega_+)}}{\lambda-\lambda_1}\underline{\bar{\psi}}_+ + \frac{i\bar{R}_{1}(-\sigma)e^{-2i(\bar{\lambda}_1x-\sigma\omega_+)}}{{\lambda}+\bar{\lambda}_1}\underline{\bar{\psi}}_-,
\end{equation}
where
\begin{equation}\label{eq3.5}
\underline{\psi}(x,\bar{\lambda}_1,\sigma) = \underline{\psi}_+\qquad \underline{\psi}(x,-\lambda_1,\sigma)=\underline{\psi}_-.
\end{equation}
Making the substitutions $\lambda = \bar{\lambda}_1$ and $\lambda=-\lambda_1$, equation (\ref{eq3.4}) gives us
\begin{equation}\label{eq3.6}
\begin{split}
\underline{\psi}_+&=1-f_+\underline{\bar{\psi}}_+-\bar{\beta}\bar{f}_-\underline{\bar{\psi}}_-,\\
\underline{\psi}_-&=1-{\beta}f_+\underline{\bar{\psi}}_+-\bar{f}_-\underline{\bar{\psi}}_-,
\end{split}
\end{equation}
with
\begin{equation}\label{eq3.7}
\begin{cases}
 f_\pm=\frac{R_1(0,\pm\sigma)}{2\nu}e^{2i(\lambda_1x\mp\sigma\lambda_1^2t)}e^{\pm2i\sigma\omega_+}\\
 \lambda_1=\mu+i\nu=le^{i\theta_1}\\
 \beta = \frac{i\nu}{\lambda_1}
\end{cases}
\end{equation}
where we use $R_1(\sigma)=R_{1}(0,\sigma)e^{-2i\sigma\lambda_1^2t}$ as shown in \cite{IL2012}.

\subsection{The discrete Jost functions}
Expressed as a matrix relationship, we see that equation \eqref{eq3.6} may be written as
\begin{equation}\label{eq3.8}
\left(
	\begin{array}{c}
		\underline{\psi}_+\\
		\underline{\psi}_-\\
	\end{array}
\right) =
			\left(
			\begin{array}{c}
				1\\
				1\\
			\end{array}
			\right) -
					 \left(
					 	\begin{array}{cc}
					 		f_+ & \bar{\beta}\bar{f}_-\\
					 		\beta f_+ & \bar{f}_-
					 	\end{array}
					 \right)
					 \left(
					 	\begin{array}{c}
					 		\underline{\bar{\psi}}_+\\
					 		\underline{\bar{\psi}}_-\\
					 	\end{array}
					 \right)
\end{equation}
To simplify the following, we define
\begin{equation}\label{eq3.9}
\underline{\bm{\Psi}} = \left(
		  \begin{array}{c}
		    \underline{\psi}_+\\
		    \underline{\psi}_-\\
		   \end{array}
	    \right), \quad
	    \mathbf{T}=\left(
			   \begin{array}{cc}
			      f_+ & \bar{\beta}\bar{f}_-\\
			       {\beta}f_+ & \bar{f}_-
			   \end{array}										  	
			\right),\quad
			\mathbf{e}=\left(
				    \begin{array}{c}
					1\\
					1\\
				    \end{array}
				   \right),
\end{equation}
in which case equation (\ref{eq3.8}) and its complex conjugate may be written as
\begin{equation}\label{eq3.10}
\begin{split}
	\underline{\bm{\Psi}} = \mathbf{e} - \mathbf{T}\cdot\bar{\bm{\Psi}}\qquad
	\underline{\bar{\bm{\Psi}}} = \mathbf{e} - \bar{\mathbf{T}}\cdot{\bm{\Psi}},
\end{split}
\end{equation}
from which we  deduce
\begin{equation}\label{eq3.11}
\bar{\underline{\mathbf{\Psi}}}=\left(\bar{\mathbf{T}}^{-1}-\mathbf{T}\right)^{-1}\cdot\left(\bar{\mathbf{T}}^{-1}-\mathds{1}\right)\cdot\mathbf{e},
\end{equation}
$\mathds{1}$ being the $2\times2$-identity matrix. 

Letting $\frac{R_1(0,\sigma)}{2\nu}=e^{x_0+i\sigma t_0}$, where $(x_0,t_0)$ are arbitrary integration constants, it follows from equation \eqref{eq3.7} that
\begin{equation}\label{eq3.12}
\begin{cases}
    f_{\pm}=\frac{l}{\mu}e^{-\xi_{\pm}}e^{i(\tilde{\theta}_{\pm}\pm2\sigma\omega_+)}\\
    \xi_\pm=2\nu(x\mp2\sigma\mu t)-x_0+\ln\left(\frac{l}{\mu}\right)\\
    \theta_{\pm}=2(\mu x\mp\sigma(\mu^2-\nu^2)t)\pm\sigma t_0.
\end{cases}
\end{equation}
We introduce a factor of $\frac{l}{\mu}$ in $f_{\pm}$, and therefore compensate by including a shift in $\xi_{\pm}$, as this will reveal greater symmetry between $\mathbf{T}$ and 
$\bar{\mathbf{T}}^{-1}$. 

Equations \eqref{eq3.7}, \eqref{eq3.9} and \eqref{eq3.12} yield
\begin{equation}\label{eq3.13}
\begin{cases}
\mathbf{T}=\frac{l}{\mu}\left(
                        \begin{matrix}
                            e^{-\xi_+}e^{i\theta_+}&\bar{\beta}e^{-\xi_-}e^{-i\theta_-}\\
                            \beta e^{-\xi_+}e^{i\theta_+}&e^{-\xi_-}e^{-i\theta_-}
                        \end{matrix}  
                        \right)e^{2i\sigma\omega_+}\\
\det(\mathbf{T})=e^{-2\xi_0}e^{i(\theta_+-\theta_-)}e^{4i\sigma\omega_+}\\
2\xi_0=\xi_++\xi_-
\end{cases}
\end{equation}
having used $1-\abs{\beta}^2=\frac{\mu^2}{l^2}$, Hence,  we observe that
\begin{equation}\label{eq3.14}
\bar{\mathbf{T}}^{-1}=\frac{l}{\mu}\left(
                        \begin{matrix}
                            e^{\xi_+}e^{i\theta_+}&-{\beta}e^{\xi_+}e^{i\theta_+}\\
                            -\bar{\beta} e^{\xi_-}e^{-i\theta_-}&e^{\xi_-}e^{-i\theta_-}
                        \end{matrix}
                        \right)e^{2i\sigma\omega_+}.
\end{equation}
and using equations \eqref{eq3.13}--\eqref{eq3.14} we then have
\begin{multline}\label{eq3.15}
\bar{\mathbf{T}}^{-1}-\mathbf{T}=\\
\frac{l}{\mu}
\left(
\begin{matrix}
2e^{i\theta_+}\sinh\xi_+&-\bar{\beta}e^{-\xi_-}e^{-i\theta_-}-\beta e^{\xi_+}e^{i\theta_+}\\
-\bar{\beta}e^{\xi_-}e^{-i\theta_-}-\beta e^{-\xi_+}e^{i\theta_+}&2e^{-i\theta_-}\sinh\xi_-
\end{matrix} \right)e^{2i\sigma\omega_+}.
\end{multline}
To obtain the inverse, we write the determinant of $\bar{\mathbf{T}}^{-1}-\mathbf{T}$ according to
\begin{equation}\label{eq3.16}
\begin{cases}
\det(\bar{\mathbf{T}}^{-1}-\mathbf{T})=2\frac{l^2}{\mu^2}\Gamma e^{i(\theta_+-\theta_-)}e^{4i\sigma\omega_+}\\
\Gamma=2\sinh\xi_+\sinh\xi_-+\frac{\nu^2}{l^2}\left(\cos2(\theta_0-\theta_1)-\cosh(2\xi_0)\right)\\
2\theta_0=\theta_++\theta_-,
\end{cases}
\end{equation}
and so the inverse is given by
\begin{multline}\label{eq3.17}
(\bar{\mathbf{T}}^{-1}-\mathbf{T})^{-1}=\\
\frac{\mu}{2l\Gamma}
\left(
\begin{matrix}
2e^{-i\theta_+}\sinh\xi_-&\bar{\beta}e^{-\xi_-}e^{-i\theta_+}+\beta e^{\xi_+}e^{i\theta_-}\\
\bar{\beta}e^{\xi_-}e^{-i\theta_+}+\beta e^{-\xi_+}e^{i\theta_-}&2e^{i\theta_-}\sinh\xi_+
\end{matrix} \right)e^{-2i\sigma\omega_+}.
\end{multline}
Given equations \eqref{eq3.11} and \eqref{eq3.17}, we now have explicit expressions for the modified Jost solutions, which are given by
\begin{multline}\label{eq3.18}
\underline{\mathbf{\Psi}}=\frac{1}{2\Gamma}\left(\begin{matrix}
						  2(1-\beta)e^{\xi_+}\sinh\xi_-+\bar{\beta}(1-\bar{\beta})e^{-2i\theta_{0}}+\beta(1-\bar{\beta})e^{2\xi_0}\\
						  2(1-\bar{\beta})e^{\xi_-}\sinh\xi_++\beta(1-\beta)e^{2i\theta_{0}}+\bar{\beta}(1-\beta)e^{2\xi_0}
						 \end{matrix}\right)\\
			    -\frac{\mu}{2l\Gamma}\left(\begin{matrix}
							2e^{-i\theta_+}\sinh\xi_-+\bar{\beta}e^{-\xi_-}e^{-i\theta_+}+\beta e^{\xi_+}e^{i\theta_-}\\
							2e^{i\theta_-}\sinh\xi_++\bar{\beta}e^{\xi_-}e^{-i\theta_+}+\beta e^{-\xi_+}e^{i\theta_-}
						       \end{matrix} \right)e^{-2i\sigma\omega_+}.
\end{multline}
\section{The Soliton Solution}
\subsection{The fluid velocity}
We note from equations (\ref{eq2.5}) and (\ref{eq2.8}) that the eigenfunction $\underline{\psi}(x,0,\sigma)$ may be written as
\begin{equation}\label{eq4.1}
\underline{\psi}(x,0,\sigma) = 1 -2\mathbf{f}\cdot\underline{\bar{\mathbf{\Psi}}},\qquad\mathbf{f}=\frac{l}{\mu}\left(\begin{matrix}\beta e^{-\xi_+}e^{i\theta_+}&\bar{\beta}e^{-\xi_-}e^{-i\theta_-}\end{matrix}\right)e^{2i\omega_+}.
\end{equation}
Using equations \eqref{eq3.18} and \eqref{eq4.1} we find
\begin{multline}\label{eq4.2}
 \underline{\psi}(x,0,\sigma)=1+\frac{2}{\Gamma}\left(\frac{\nu^2}{l^2}(\cosh2\Xi-\cos2(\theta_0-\theta_1))+i\frac{\mu\nu}{l^2}\sinh2\Xi\right)\\
 -i\frac{2}{\Gamma}\frac{\mu\nu}{l^2}\left(\sinh\xi_-e^{i(\theta_+-\theta_1)}-\sinh\xi_+e^{-i(\theta_--\theta_1)}\right)e^{2i\sigma\omega_+}
\end{multline}
with $2\Xi=\xi_--\xi_+$. 

The expression \eqref{eq4.2} may be  used to compute  the velocity component $u(x,t)$ via the relation
\begin{equation}\label{eq4.4}
	e^{2i\sigma\omega_+} = \frac{\underline{\psi}(x,0,\sigma)}{\underline{\bar{\psi}}(x,0,\sigma)}\Rightarrow e^{2i\sigma\omega_+}=\frac{N(x,\sigma)}{\bar{N}(x,\sigma)}
\end{equation}
(see \cite{IL2012} for further discussion), where we introduce
\begin{multline}\label{eq4.5}
 N(x,\sigma)=1+\frac{2}{\Gamma}\left(\frac{\nu^2}{l^2}(\cosh2\Xi-\cos2(\theta_0-\theta_1))+i\frac{\mu\nu}{l^2}\sinh2\Xi\right)\\
 -i\frac{2}{\Gamma}\frac{\mu\nu}{l^2}\left(\sinh\xi_-e^{-i(\theta_+-\theta_1)}-\sinh\xi_+e^{i(\theta_--\theta_1)}\right).
\end{multline}
Referring the equation \eqref{eq2.5} we observe that $\omega_+$ may be written according to
\begin{equation}\label{eq4.6}
\omega_{+}:=\frac{1}{2}\int_{x}^{\infty}u(\xi,t)\mathrm{d}\xi =  -\frac{i\sigma}{2}\ln\left(\frac{N(x,\sigma)}{\bar{N}(x,\sigma)}\right),
\end{equation}
and upon applying the fundamental theorem of calculus we obtain
\begin{equation}\label{eq4.7}
u= i\sigma\frac{\bar{N}(x,\sigma)N_x(x,\sigma)-N(x,\sigma)\bar{N}_{x}(x,\sigma)}{N(x,\sigma)\bar{N}(x,\sigma)}
\end{equation}
thus ensuring the fluid velocity $u(x,t)$ may be written in terms of the scattering data.

\begin{remark}
It is easily deduced from equation \eqref{eq4.5} that $N(x,\sigma) = \bar{N}(x,-\sigma)$, thereby ensuring $u(x,t)$ is real and independent of $\sigma$, as expected.
\end{remark}

\subsection{The surface elevation}
In \cite{IL2012} it was found that as $\lambda\to\infty$ then the discrete Jost functions may be expanded as follows:
\begin{equation}\label{eq4.8}
\underline{\psi}(\lambda,\sigma)=1-\lambda^{-1}\left[\frac{\sigma u}{4}+\frac{i}{8}\int_{x}^{\infty}\left(u^2+4w\right)dy\right] +\mathcal{O}(\lambda^{-2}).
\end{equation}
Referring to  equation \eqref{eq2.1} where the auxiliary potential $w$ is introduced, and noting that the spectral problem corresponds to the system \eqref{KB} when $\kappa=-\frac{1}{2}$, it follows that the integrand in equation \eqref{eq4.8} is given by
\begin{equation*}\label{eq4.9}
    u^2+4w = (1+2\kappa)u^2+4\eta=\eta.
\end{equation*}
Hence for $\kappa=-\frac{1}{2}$, equation \eqref{eq4.8} becomes
\begin{equation}\label{eq4.11}
\underline{\psi}(\lambda,\sigma)=1-\frac{1}{\lambda}\left[\frac{\sigma u}{4}+\frac{i}{2}\int_{x}^{\infty}\eta dy\right]+\mathcal{O}(\lambda^{-2}).
\end{equation}
Using this form of $\underline{\psi}(x,\lambda,\sigma)$, we may now construct the surface elevation $\eta$ in terms of the scattering data.

Asymptotically, we also have
\[\frac{2i\nu}{\lambda-\lambda_i}\to\frac{2i\nu}{\lambda}\quad \text{ as }\lambda\to\infty,\]
in which case we have
\begin{equation}\label{eq4.12}
\begin{split}
\underline{\psi}(\lambda,\sigma) &\to 1 + \frac{2i\nu}{\lambda}\left[f_+\underline{\bar{\psi}}_+
+\bar{f}_-\underline{\bar{\psi}}_-\right]. \text{as }\lambda\to\infty.
\end{split}
\end{equation}
We let $\mathbf{g} = \left(f_+\quad f_-\right)$ and compare terms of order $\mathcal{O}(\lambda^{-1}),$ from which we obtain
\begin{equation}\label{eq4.13}
	-\frac{\sigma u}{4}-\frac{i}{8}\int_{x}^{\infty}\eta dy=2i\nu\mathbf{g}\cdot\underline{\Psi}
\end{equation}
and subtracting the complex conjugate we find
\begin{equation}\label{eq4.15}
\begin{aligned}
	-\frac{i}{4}\int_{x}^{\infty}\eta dy&=2i\nu\left(\mathbf{g}\cdot\underline{\bar{\Psi}}+\bar{\mathbf{g}}\cdot\underline{{\Psi}}\right)
	\Rightarrow \eta=8\nu\left(\mathbf{g}\cdot\underline{\bar{\Psi}}+\bar{\mathbf{g}}\cdot\underline{{\Psi}}\right)_{x}
\end{aligned}
\end{equation}
upon applying the fundamental theorem of calculus.

Referring to equation \eqref{eq3.18}, we have
\begin{multline}\label{eq4.16}
\mathbf{g}\cdot\mathbf{\Psi}=1-\frac{1}{\Gamma}\left(\frac{\mu^2}{l^2}\sinh2\xi_0-\frac{\mu\nu}{l^2}\sin2\tilde{\phi}_0\right)\\
+\frac{\mu^2}{l^2\Gamma}\left(\sinh\xi_-e^{-\tilde{\phi}_+}+\sinh\xi_+e^{-\tilde{\phi}_-}\right)e^{2i\sigma\omega_+}\\
+i\frac{\mu\nu}{l^2\Gamma}\left(\cosh{\xi}_-e^{-\tilde{\phi}_+}-\cosh\xi_+e^{-i\tilde{\phi}_{-}}\right)e^{2i\sigma\omega_+},
\end{multline}
and upon replacing equations \eqref{eq4.4}--\eqref{eq4.5} in equation \eqref{eq4.16}, we also find
\begin{equation}\label{eq4.17}
\begin{aligned}
\mathbf{g}\cdot\bar{\mathbf{\Psi}}&=\frac{1}{\bar{N}(\sigma)}\frac{\mu^2}{l^2}\left(\sinh\xi_-e^{i\tilde{\phi}_+}+\sinh\xi_+e^{i\tilde{\phi}_-}-\sinh(2\xi_0)\right)\\
				  &\qquad-\frac{1}{\bar{N}(\sigma)}\frac{\mu\nu}{l^2}\left(\sin(2\tilde{\phi_0})+i\cosh\xi_-e^{i\tilde{\phi}_+}-i\cosh\xi_+e^{i\tilde{\phi}_-}\right)\\
				  &=-\frac{1}{4\nu}\left(\ln(\bar{N}(\sigma))\right)_x.
\end{aligned}
\end{equation}
It follows from equations  \eqref{eq4.15} and \eqref{eq4.17} that
\begin{equation}\label{eq4.18}
 \eta(x,t)=-2\ln\left(N(\sigma)\bar{N}(\sigma)\right)_{xx},
\end{equation}
thus giving us an explicit expression for the surface elevation $\eta$ in terms of the discrete scattering data $\{\lambda_1, R_1(\sigma)\}$.

To remove all extra phase shifts, we use the fact the spectral problems \eqref{eq2.1} are translation invariant under coordinate transformations of the form
\[x\to x+x_1,\quad x_1\in\RR.\]
Imposing such a coordinate transformation, we now choose, $x_0$, $x_1$ and $t_0$ as follows:
\begin{equation}\label{eq4.3}
\begin{cases}
2\nu x_1-x_0+\ln\left(\frac{\mu}{l}\right)=0\\
2\mu x_1\pm t_0-\theta_1=0\\
\Rightarrow t_0=0\quad x_1=\frac{1}{2\mu}\theta_1\quad x_0=\frac{\nu}{\mu}\theta_1+\ln\left(\frac{l}{\mu}\right).
\end{cases}
\end{equation}
Choosing $(x_0,t_0)$ in this way essentially means we hav chosen our intial data $R_1(0,\sigma)$ and correspondingly $\left(u(x,0),\eta(x,0)\right)$ to ensure those values given in equation \eqref{eq4.3}.

\section{conclustion}
\begin{figure}[h!]
\includegraphics[width=0.35\textwidth]{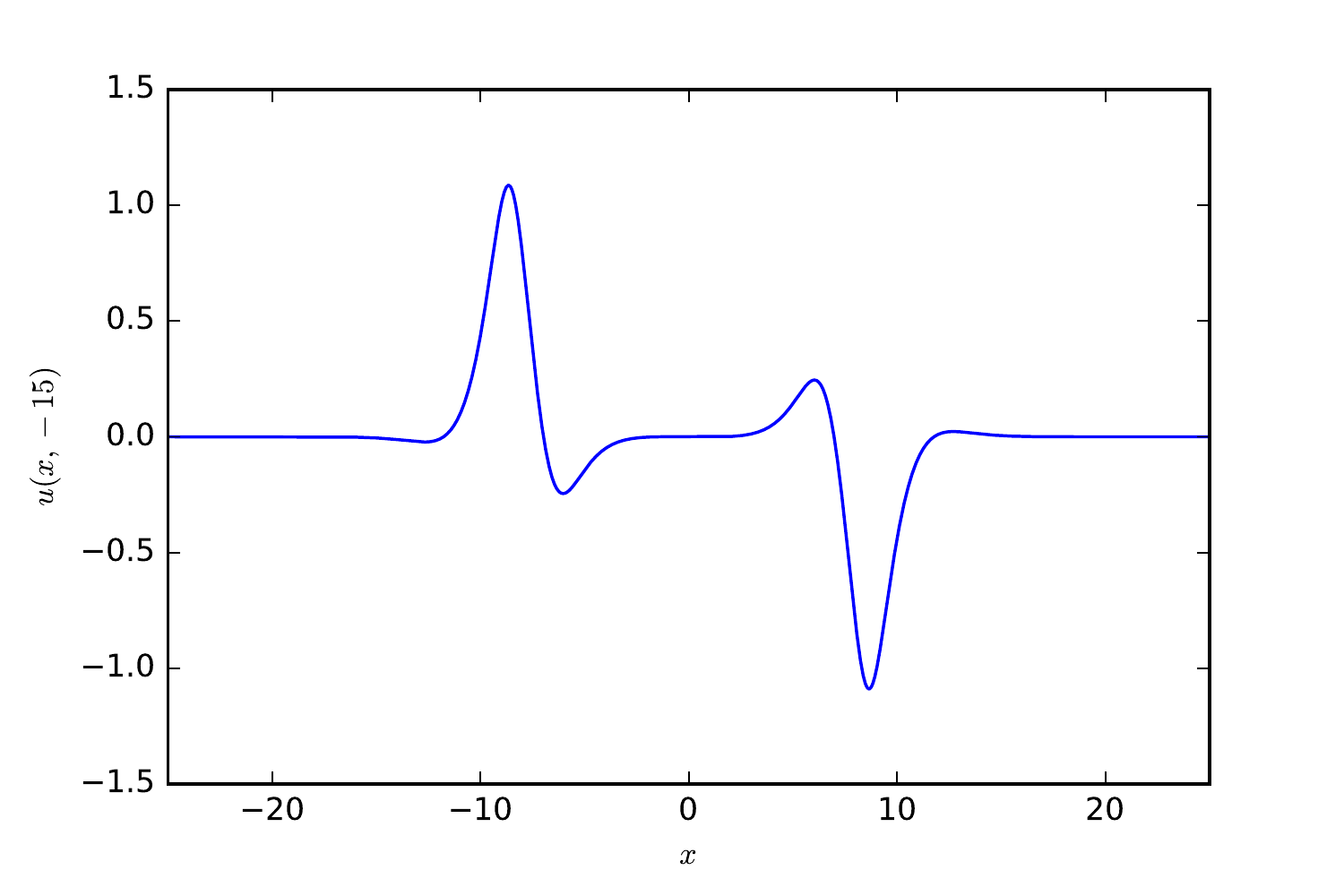}\includegraphics[width=0.35\textwidth]{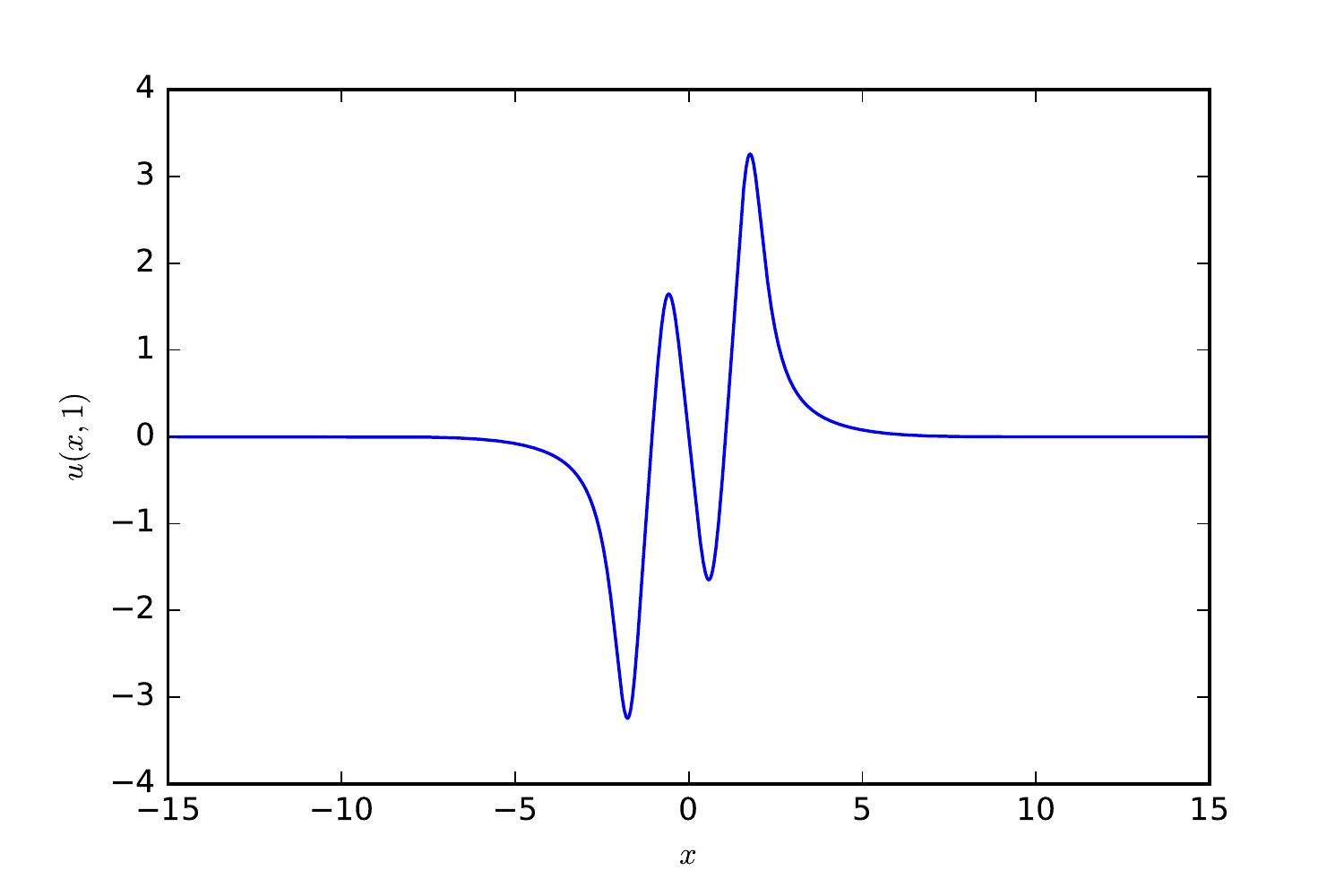}\includegraphics[width=0.35\textwidth]{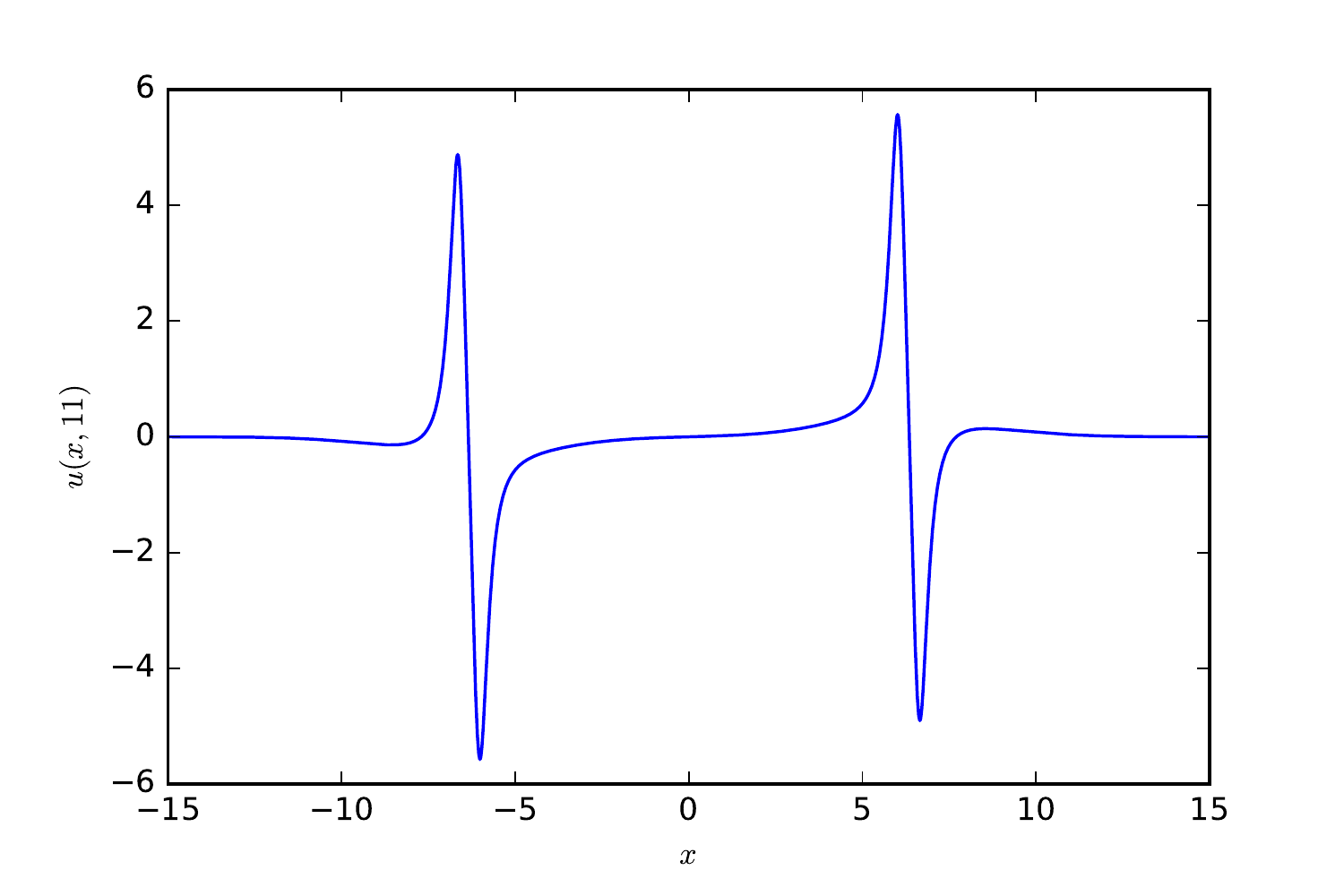}
\includegraphics[width=0.35\textwidth]{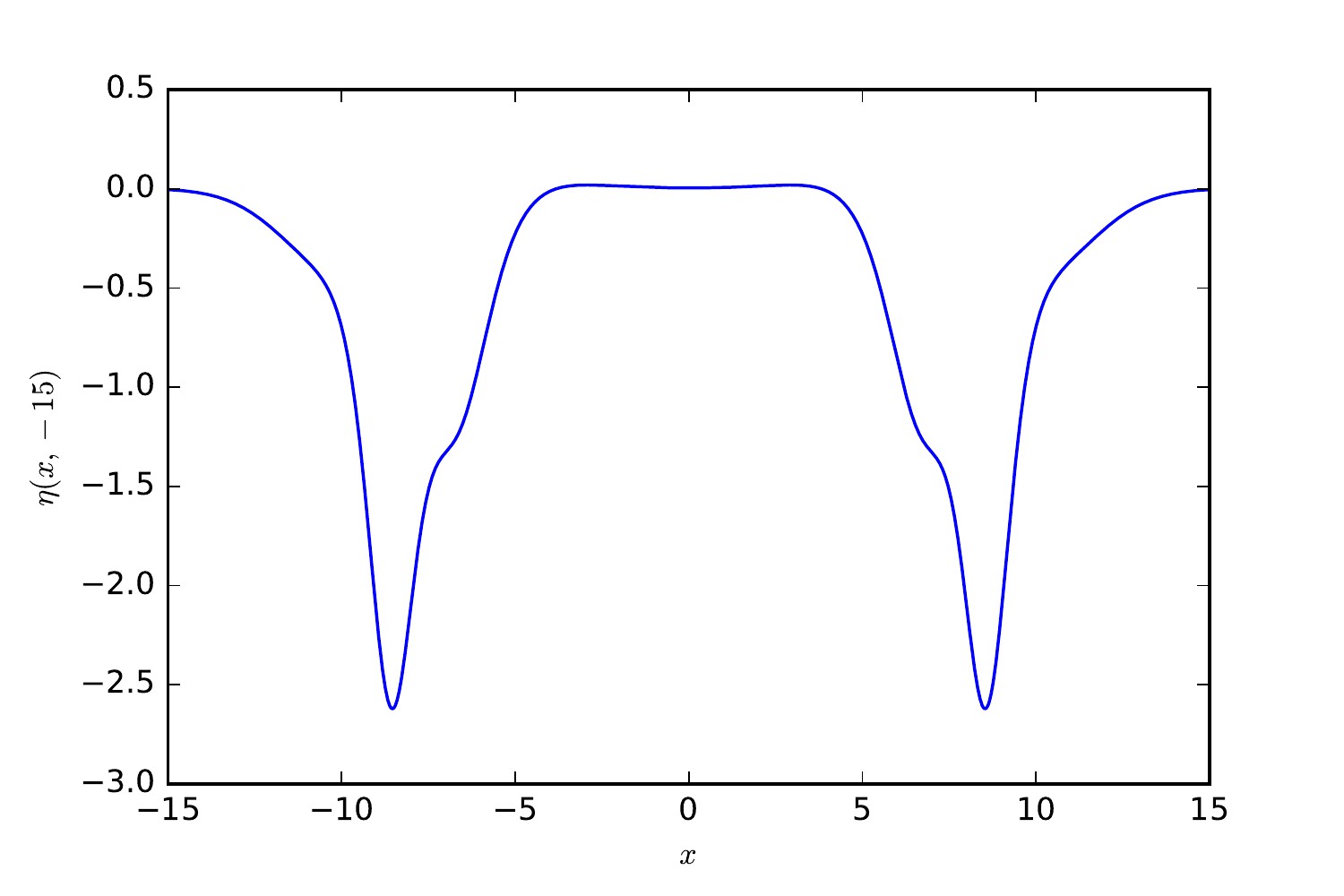}\includegraphics[width=0.35\textwidth]{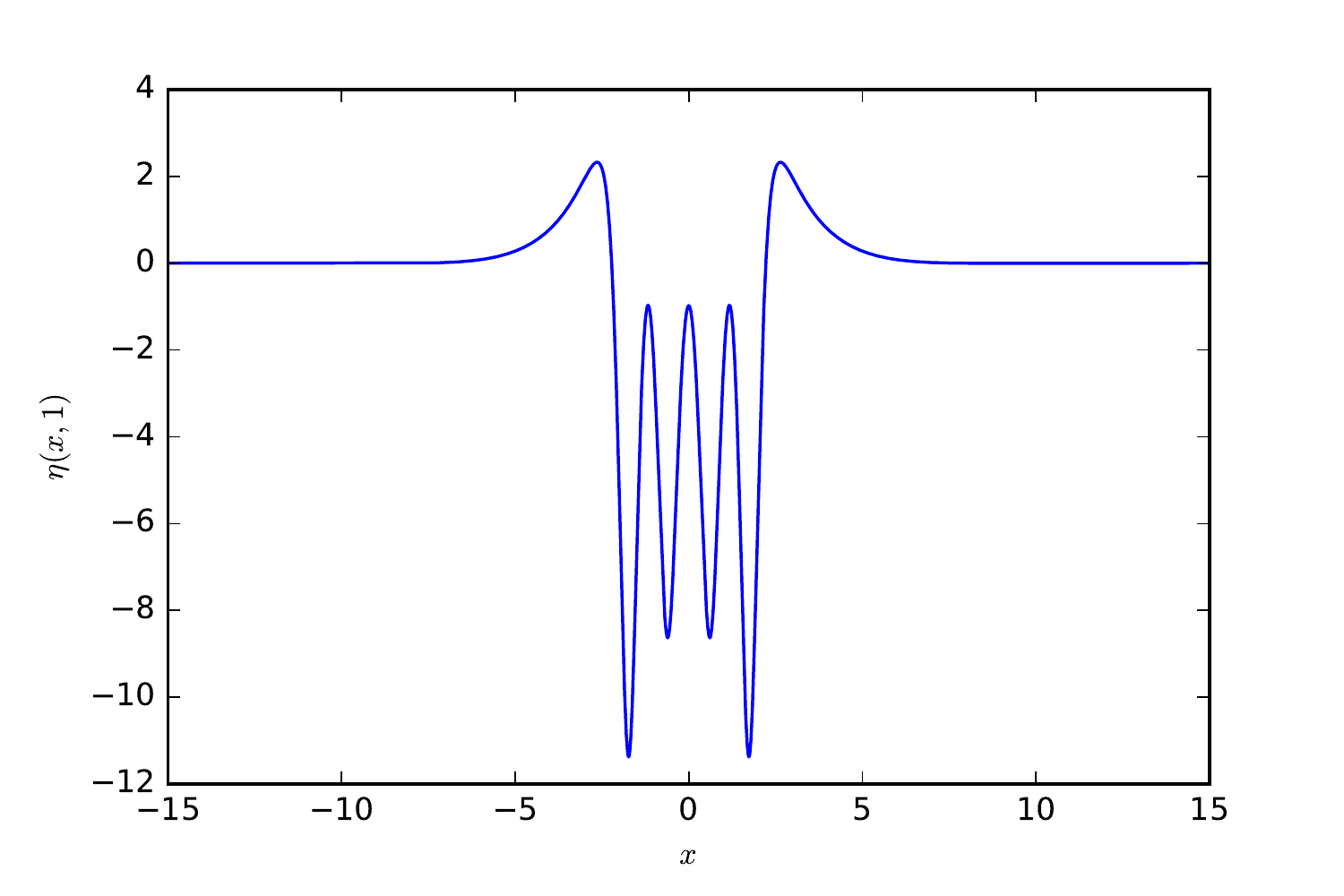}\includegraphics[width=0.35\textwidth]{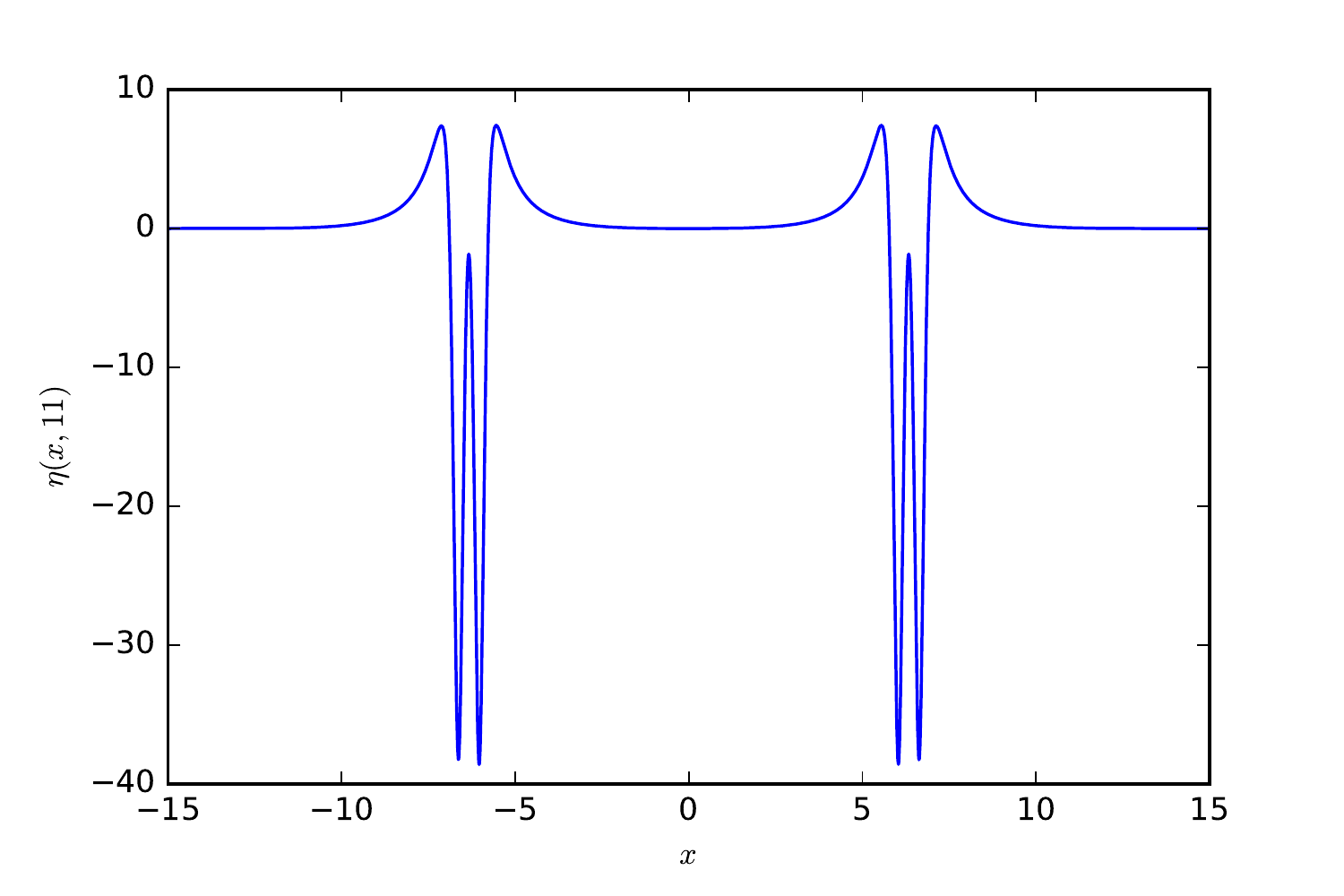}
\caption{Snapshot of the soliton solution of the sytem \eqref{KB} with discrete spectral parameter $\lambda_1=0.25+0.5i$ at times $t=-15$, $t=1$ and $t=11$, running from left to right. The upper panel show the velocity $u(x,t)$ and the lower panel shows the surface elevation $\eta(x,t)$.}\label{fig2}
\end{figure}

The soliton solution for the system \eqref{KB} we obtain is illustrated in Figure \ref{fig2}. The solution we illustrate corresponds to the discrete spectrum $\lambda_1=0.25+0.5i$. The form of this solution is in line with expectations from that obtained in \cite{IL2012}, displaying the same symmetry about the line $x=0$ at all times $t$. Moreover, the soliton solution we obtain here displays periodic singularities, in line with those displayed by the topological solution. It appears the soliton obtained here is non-physical, and is a reflection of the fact that  travelling wave solutions of the Euler equation in the absence of surface tension are unstable. 

\begin{figure}[h!]
\includegraphics[width=0.5\textwidth]{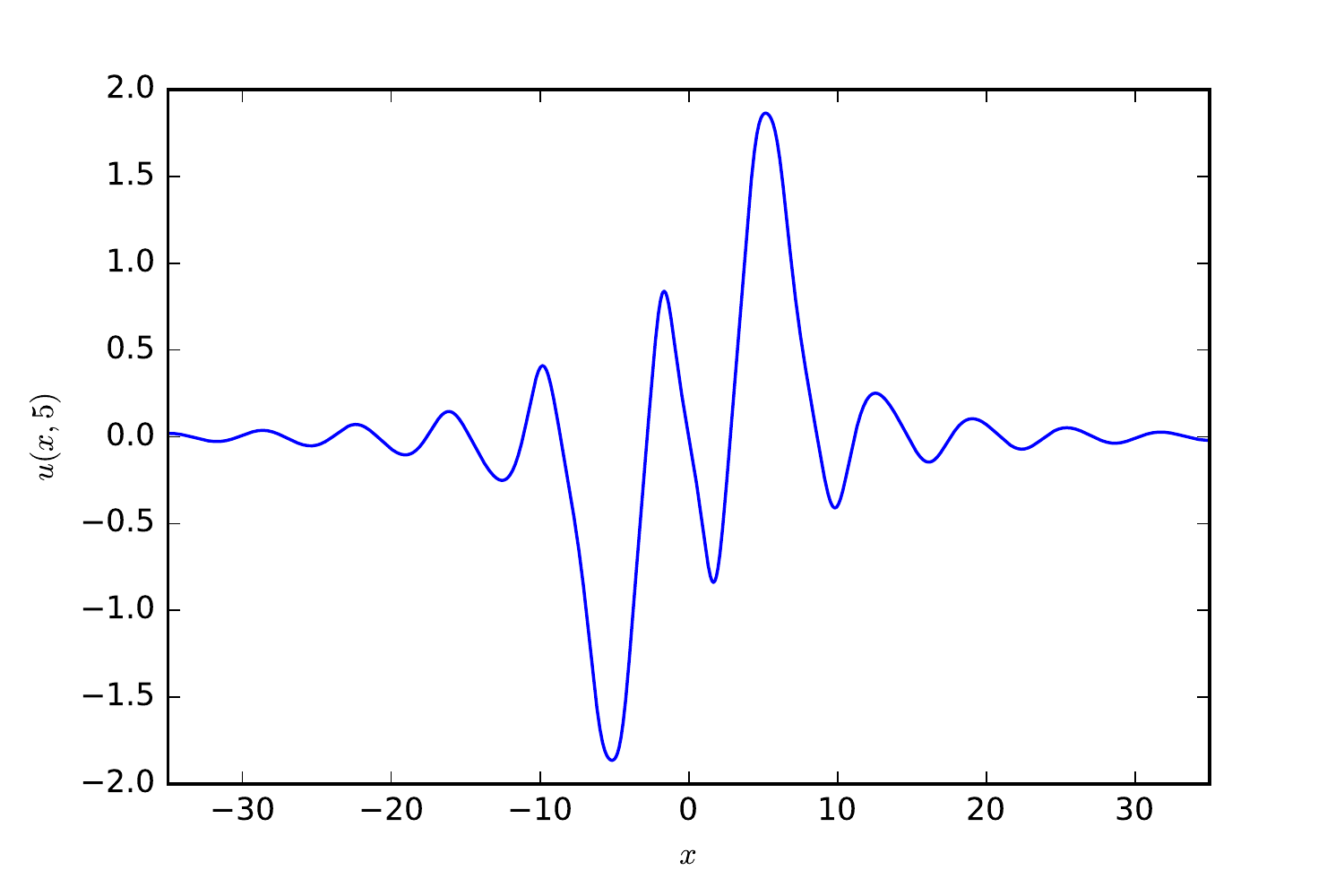}\includegraphics[width=0.5\textwidth]{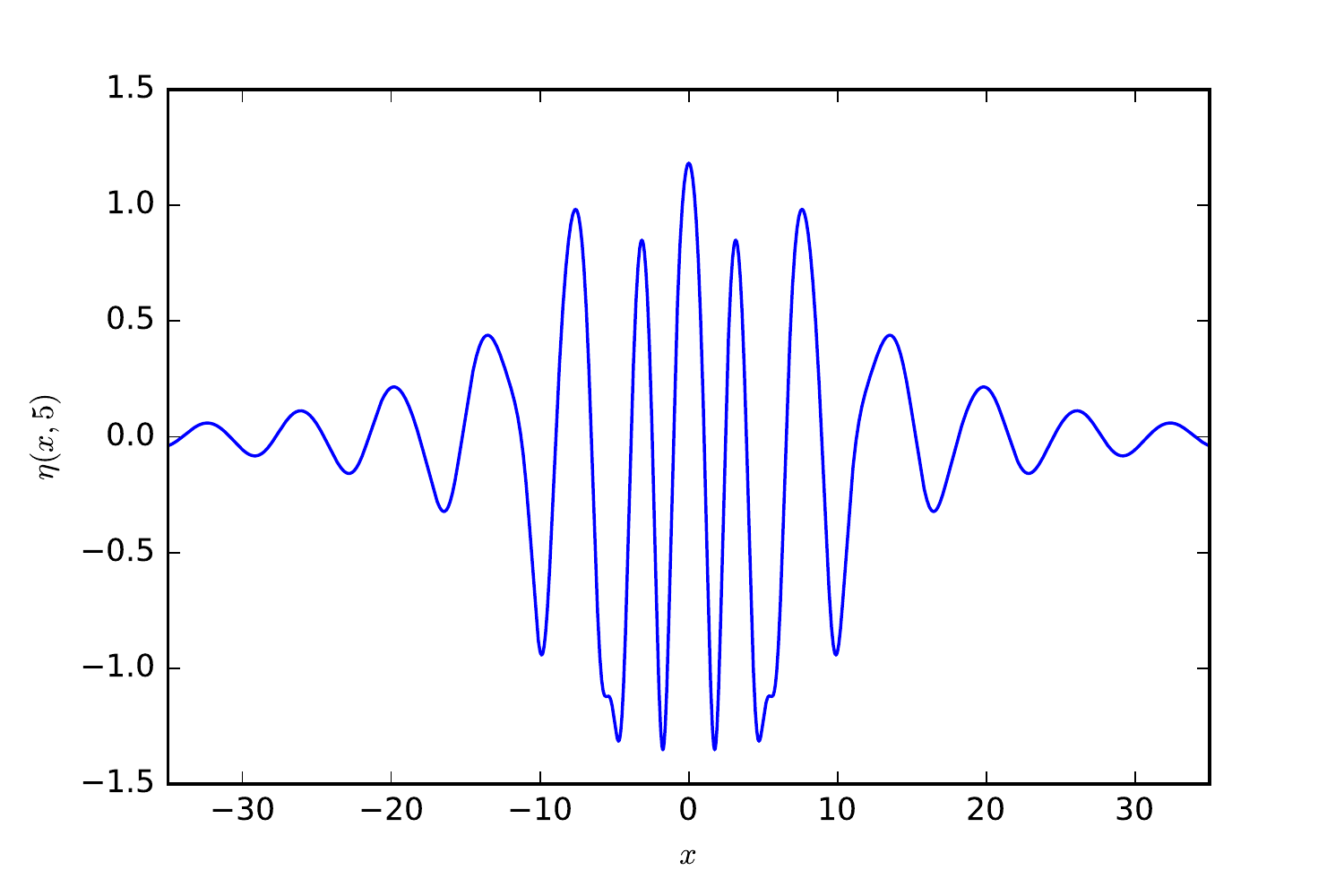}
\caption{The soliton solution of the system \ref{KB} for $\lambda_1=0.5+0.05i$ at time $t=5$. The left-hand panel shows the velocity $u(x,t)$ while the right-hand one show the surface elevation $\eta(x,t)$. }\label{fig3}
\end{figure}

In \cite{IL2012} it was expected that the hydrodynamical relevance of the system  related to solutions along its continuous spectrum, which should be unstable due to the lack of surface tension in the model. While we have set $r(\lambda,\sigma)=0$ here, as $\lambda_1$ approaches the real axis the solution appears to display disperive behaviour. To illustrate this, in Figure \ref{fig3} we include a plot of the solution $\{u,\eta\}$ when $\lambda_1=0.5+0.05\nu$, and indeed similar behaviour was observed in all solutions we plotted with $\mu>\nu.$    

\section{Acknowledgements}
TL is grateful to Rossen Ivanov for numerous helpful discussions.

\end{document}